\documentclass[twocolumn,aps,prb,showpacs,reprint,showkeys]{revtex4-1}

\usepackage{times}

\usepackage{amsmath,amssymb}           

\usepackage{graphicx}                   
\usepackage{subfigure}                  

\DeclareTextSymbol{\degre}{OT1}{23}

\begin{document}

\title{Far field subwavelength imaging and focusing using a wire medium based resonant metalens}
\author{Fabrice Lemoult}
\email{fabrice.lemoult@espci.fr}
\author{Geoffroy Lerosey}
\author{Mathias Fink}
\affiliation{Institut Langevin, ESPCI ParisTech \& CNRS, 10 rue Vauquelin, 75231 Paris Cedex 05, France}

\date{\today}

\begin{abstract}
This is the second article in a series of two dealing with the concept of "resonant metalens" we introduced recently [Phys. Rev. Lett. 104, 203901 (2010)]. It is a  new  type  of  lens capable  of  coding  in  time  and  radiating  efficiently  in  the  far field region sub-diffraction information of an object. A proof of concept of such a lens is performed in the microwave range, using a medium made out of a square lattice of parallel conducting wires with finite length. We investigate a sub-wavelength focusing scheme with time reversal and demonstrate experimentally spots with focal widths of $\lambda/25$. Through a cross-correlation based imaging procedure we show an image reconstruction with a resolution of $\lambda/80$. Eventually we discuss the limitations of such a lens which reside essentially in losses.
\end{abstract}


\maketitle

\section{Introduction \label{sample}}
Conventional imaging devices, such as optical lenses, create images by capturing the waves emitted by an object and then bending them. The image obtained is diffraction limited by the Rayleigh criterion at best equal to half the operating wavelength. Indeed, the information of the object on a smaller scale is carried by evanescent waves that decrease exponentially from the surface of the object:  these waves  never reach the image plane explaining the Rayleigh limit.

In order to increase the image resolution, the first idea that arised was to measure the evanescent field: Synge \cite{Synge} proposed in the early 20th century the first near field imaging procedure. This first work has lead to the development of various near field scanning microscopes based on the measurement of the evanescent wavefield\cite{ash,lewis,pohl}. These methods require multiple scans of the object and they cannot be used for living cells that requires real-time imaging.

In 2000, J. Pendry \cite{Pendry} suggested that a material with a negative index of refraction \cite{Veselago} could make a perfect lens. Such a "superlens" has first interested the microwaves community, until it reached the optical range in 2005, when Zhang's group \cite{zhang} verified experimentally that optical evanescent waves could indeed be enhanced as they passed through a silver superlens. They imaged objects as small as $90 \textrm{ nm}$  with their superlens, five times the resolution limit of classical optical devices. Nevertheless, even if the "superlens" creates a perfect image the need of a near field scan is still required. 

There is obviously a need of far field methods for imaging living cells in real time. The most recent ideas for perfect imaging reside in a projection of the near field profile by the use of a diffraction grating \cite{liu} or by a first step of magnification before conventional imaging \cite{narimanov,engheta}. In the former case the resolution is limited by the short range of wave numbers accessible in the far field, and in the latter by the geometry of the magnifying lens. Moreover, the intrinsic losses of real materials diminish the resolution ability of these devices. Nowadays, many works concentrate on surface plasmons polaritons since their dispersion relation allows propagation of waves with high wave numbers that carry informations smaller than the operating wavelength.
 	
In a recent letter \cite{metalens}, we proposed the concept of a resonant metalens: a cluster of strongly coupled resonators illuminated with broadband wave fields. The concept is the following: the small details of the object, usually carried by evanescent waves, are in this case converted onto propagating ones: measuring the far field one can obtain sub-wavelength  information about the object. We achieved far field imaging and focusing experiments with resolutions of respectively  $\lambda/80$ and  $\lambda/25$, well below the diffraction limit. Like the initial development of the Pendry's "superlens", we performed a proof of concept in the microwave range because of experimental easiness but the transposition toward optical range is not speculative since sub-wavelength optical resonators are designable. 

We focus on conducting wires with finite length used as resonators, and we scope on the resonant metalens introduced initially \cite{metalens} consisting of a square lattice of $N\times N$ ($N=20$) identical metallic wires, aligned along the vertical axis (defining the longitudinal $z$-direction). The length of the wires along $z$ is equal to $40 \textrm{ cm}$ and their diameter to $3\textrm{ mm}$. The period lattice in both transverse directions ($xy$ plane) is $1.2 \textrm{ cm}$, which is roughly equals to $\lambda/70$ for the first intrinsic resonance of a single wire. The experimental sample is made out of copper wires and a Teflon structure.

\begin{figure}
\includegraphics[width=\columnwidth]{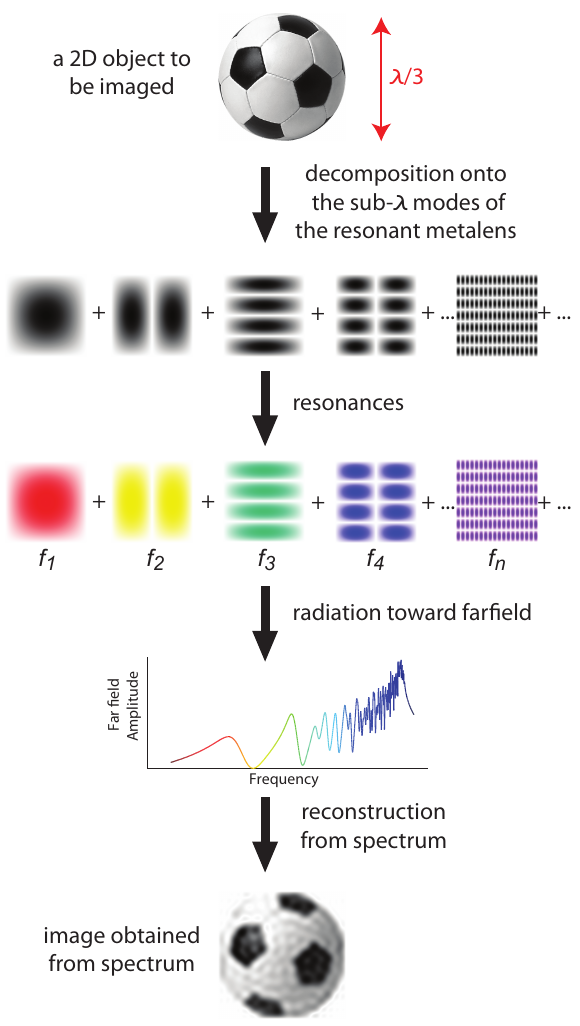}
\caption{\label{frequency_coding} (color online) Schematic representation of the resonant metalens mechanism. A sub-wavelength object is placed at the input of the resonant metalens. The broadband near field emitted by the object decomposes onto the sub-wavelength eigenmodes supported by the metalens. Each mode experiments its own resonance and radiates field toward far-field at its own resonance frequency: the resonant metalens has colored the emission. Then by exploiting the dispersion relation and the radiated spectrum, a super-resolved image can be built. The resolution of the image will be discussed in the last part of this article.}
\end{figure}

During the first article we have demonstrated that a sub-wavelength broadband source placed at the input of the metalens excites the sub-wavelength varying eigenmodes therein. Then we have shown that each mode radiates in the far field, and thanks to dispersion, the sub-wavelength information of the source reaches the far field region stored in the spectrum. In the figure \ref{frequency_coding} we sketch the physical mechanisms that permit the super-resolution properties of this system. The radiated field contains the details of the object as a frequency signature, and exploiting this signature with the dispersion relation permits to build an image. We will show how one can exploit these phenomena to achieve sub-wavelength imaging/focusing from the far field. A time reversal scheme for sub-wavelength focusing is presented. The reciprocal counterpart of focusing, the imaging is then presented. The imaging procedure is a derivative of time reversal because it consists in cross-correlation imaging with knowledge of the initial Green's functions. Finally, we discuss the impact of losses inherent to experimental devices and link them to the obtainable resolution of our device.

\section{Time Reversal subwavelength focusing}

The resonant metalens radiates fields that contain the temporal signature of the near field profile, and the step is now the exploitation of this temporal code for focusing on a deep subwavelength scale. Based on works made at the laboratory, we have proposed to use the time reversal technique to exploit the radiation. Time reversal is a good candidate because it uses broadband waves and it has  been  proved  to  be  a powerful  principle for  the  study  of  waves  in  complex media  \cite{Derode,PRLLerosey2004},  their  use  for  focusing  \cite{Fink,Lerosey2006}, or even telecommunication \cite{Henty,LeroseyRS,Montaldo2004}.

\subsection{Theory}
In a typical time reversal scheme, a dipole source emits a short pulse. The wave field propagates and is recorded with one antenna or a set of antennas, usually referred to as the time reversal mirror (TRM). Second, the recorded signals are digitized, flipped in time and transmitted back by the same set of antennas. The resulting wave is found to converge back to the initial source. 

For a narrowband signal of oscillating pulsation $\omega$, the time reversal focusing on position $\mathbf{r_0}$ is equivalent to phase conjugation, and the electric field generated at position $\mathbf{r}$ simply writes \cite{Fink2,Carminati}:

\begin{equation}
\mathbf{E}_{rt}(\mathbf{r},\omega)=-2i\mu_0\omega^2 \textrm{Im } \tensor{\mathbf{G}}(\mathbf{r},\mathbf{r_0},\omega)\mathbf{p^*}
\end{equation}
  
\noindent Where $\tensor{\mathbf{G}}$ stands for the dyadic green function and $\mathbf{p}$ represents the initial dipole source. When the dyadic green function is the free space one, it results in a cardinal sine function: the focal spot is diffraction limited to $\lambda/2$. When the time reversal mirror is placed in the near field of the initial source the green function takes into account the evanescent component of the field generated by the source and a smaller spot can be obtained \cite{deRosny}.

Note that the field amplitude at the focal point is proportional to the imaginary part of the Green's function which is itself proportional to the so-called local density of states\cite{McPhedran} (LDOS). In the special case of the resonant metalens presented here, the LDOS depends on the TEM Bloch modes that resonate at the operating angular frequency $\omega$. However, for broadband excitation, the resulting field takes advantage of the frequency diversity. For an excitation with a flat bandwidth $\Delta \omega$, all frequencies add up in phase at a given time, the collapse time (t = 0). At this specific time, the time reversed field writes \cite{Fink2}:

\begin{equation}
 \mathbf{E}_{rt}(\mathbf{r},t=0)\propto \mu_0\omega^2 \int_{\Delta\omega}\textrm{Im } \tensor{\mathbf{G}}(\mathbf{r},\mathbf{r_0},\omega)\mathbf{p^*}\textrm{d}\omega
 \end{equation} 
 
Thus, the time reversed field at the source point and at the focal time is directly proportional to the number of independent TEM Bloch modes excited by the source inside the metalens. In the presence of the metalens, the near field of the waves inside it allows the Green's tensor to fluctuate on a scale smaller than the wavelength, by the way of the TEM Bloch modes, and therefore to reach a sub-wavelength focusing with time reversal \cite{deRosny}.

\begin{figure*}
\includegraphics[width=\textwidth]{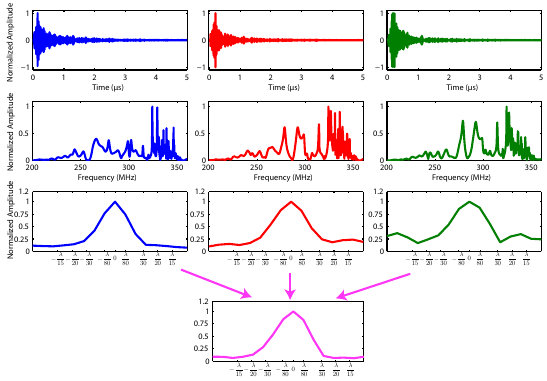}
\caption{\label{spots} Signals (top) and  spectra (middle) received in the far field after emission of a 10 ns pulse from the central monopole with the lens for distinct position of the far field receiving antenna. From left to right it corresponds respectively to an azimuthal angle of 0\degre, 45\degre and 90\degre. (bottom) The focal spot obtained for each one-channel time reversal experiment. $\lambda/25$ widths  are demonstrated in the presence of the resonant metalens for each angle. Taking advantage of the spatial diversity by doing multi-channels time reversal reduces the spatial side lobes as shown on the bottom.}
\end{figure*}

\subsection{Experiment}
All the measurements are realized in an anechoic chamber, in order to isolate the response of the resonant lens from the one of the host medium, namely the cavity used in \cite{Lerosey2007} or at the laboratory. The experimental resonant metalens, is placed vertically on a 1 by 1 meter ground plane in order to screen the cable that would otherwise induce parasitic effects. This is a convenient way to realize a near field experiment with microwaves. Another approach would have consisted in using a wax or glass prism used in total internal reflection in order to illuminate a small dipole at the input of the lens, but given the operating wavelength here, we did not choose it for practical reasons.

In the time reversal experiment, we use two 8 channels multiplexers between the generator and the sources. In this configuration, 16 small identical monopoles are placed between the ground plane and the lens, and linked to the multiplexers through the ground plane acting as a shield for the cables, by means of soldered connectors. The monopoles are linearly placed, and the separation between two consecutive monopoles is $1.2\textrm{ cm}$, ie. one period of the wire array.

We have a single receiving antenna, that will then be our time reversal mirror, namely it will be the source that focuses the waves at the output of the resonant metalens, back to the original source. First, we emit a $5 \textrm{ ns}$ pulse using one of the middle monopoles from the 16 monopoles array. Then, we record the field generated by this source using the vertically polarized antenna 6 wavelengths apart in the anechoic chamber. The signal is then digitized at a sampling rate of $10 \textrm{ GHz}$ using an oscilloscope, and flipped in time using Matlab and a computer connected to the oscilloscope. We now exchange the cables, and use the receiving antenna as a transmitting antenna, acting as a 1 channel Time Reversal mirror. The time reversed signal is sent to an arbitrary waveform generator, and used for the emission from the vertically polarized antenna.

After propagation in the anechoic chamber, the signal arrives on the resonant metalens, and we record the fields generated on the 16 monopoles, which are now receivers, switching from one monopole to the others using multiplexers. We acquire the fields on other positions using the oscilloscope and store the 16 temporal waveforms. The focal spot obtained using time reversal is then defined as the maximum of the square amplitude of each waveform across the temporal window (Fig. \ref{spots}). First, it confirms the theoretical explanation of the wire medium since a short pulse transforms onto a signal that extends in time due to resonances. The spectra also show many distinct resonance peaks, each one corresponding to its own sub-wavelength scale as stated by the theoretical dispersion relation. With a single isotropic antenna placed in free space, a sub-wavelength spot of width $\lambda/25$ is obtained at the input of the wire medium after time reversal. Here, we show that this experiment is formally the same, but simplified, as the one conducted initially \cite{Lerosey2007}.

To demonstrate the spatial diversity of the radiation, we have measured the fields produced by the structure in 8 distinct directions, tilting the ground plane and the wire medium of 45\degre each time. We show in Figure \ref{spots} the results for three measurements taken at 0\degre, 45\degre and 90\degre in the far field. The temporal signals as well as their Fourier transforms received in the far field are plotted.  The result of the same one channel time reversal scheme has been experimented for the three angles. One can notice that the 3 spectra show resonance peaks at distinct positions revealing the spatial diversity. For the scope of imaging, this property will permit to fight against modes degeneracy:  at a given frequency where multiple Bloch modes resonate it is actually possible to recover the weight of each one thanks to the spatial degrees of freedom. Eventually, we performed a three channels time reversal experiment. The focal spot obtained has the same width as previous ones, but the side lobes obtained have a lower amplitude.  

\section{Subwavelength imaging}
\begin{figure}
\includegraphics[width=\columnwidth]{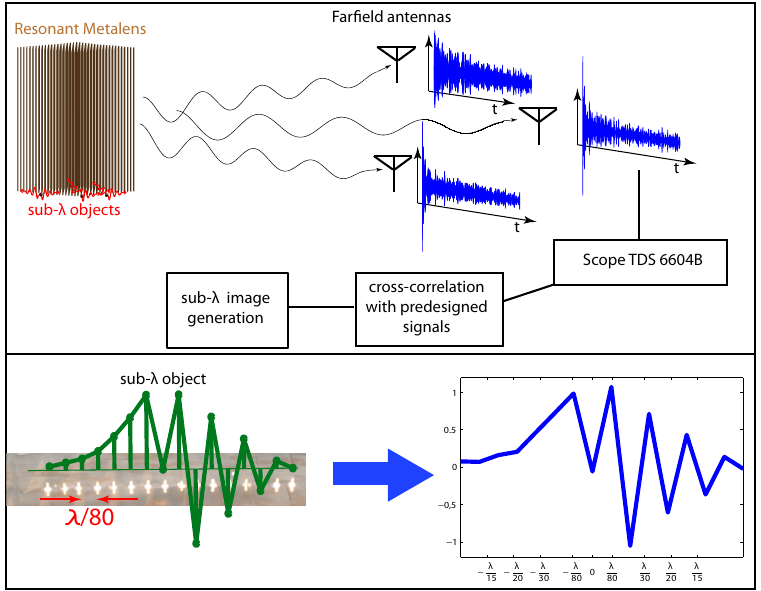}
\caption{\label{imagery} (top) Schematic representation of the subwavelength imaging procedure.  When a subwavelength source, placed near the resonant metalens, emits a broadband pulse, the radiated field is recorded with 8 antennas placed in distinct azimuthal directions. With an a priori knowledge of all of the impulse responses, a bank of filters were previously built. The image is reconstructed by cross correlation between the received signals and the filters. (bottom) The experimental sub-wavelength object (left) consists in a simultaneous emission of the 16 small monopoles (shown on picture) with a given amplitude profile (green line). An image reconstruction (right) with a resolution of $\lambda/80$ can be achieved.}
\end{figure}

With the time reversal experiment, the focusing feasibility on a sub-wavelength scale has been demonstrated. In this part, we consider the reciprocal operation, namely the imaging. In most of broadband imaging techniques, images are created by applying filters to the received RF data. The simplest filter is the delay line filter known as beamforming  that only considers free space propagation of short pulses. Its resolution is given by the Rayleigh criterion $\delta r=1.22 \lambda F/D$, where $F$ is the focal length and $D$ corresponds to the aperture length. When the propagation occurs in complex media the Green's function becomes more complicated, the beamforming shows its limitations and more complicated filters are required. The imaging procedure consists in cross-correlations between these filters and a set of received signals that supposedly carry an image information (Fig. \ref{imagery}).

As in the time reversal scheme, we propose to use filters that exploit the whole duration of the Green's functions, and a calibration step will consist in recording all of the Green's functions. Typically, with $N$ receiving antennas placed in the far field region, we first record a set of $L\times N$ Green's functions $g_{ij}(t)$. With this knowledge, we are presumably able to get $L$ distinct pixels in the image plane. The time reversed Green's functions could have been used to build the image, but it is not the best signals in terms of resolution since time reversal is an operation that maximizes the energy, and not minimizing the side lobes level.

 The idea is thus to design signals that are more like an inverse filter which has better resolution capabilities \cite{Tanter}. Inversion procedure suffers from matrix inversion and noise, and at the laboratory inversion procedures based on iterations of time reversal have been proposed \cite{Montaldo,Montaldo2004,Lemoult}. Using the signals $g_{ij}(t)$, we then construct a bank of $L \times N$ pseudo-inverse signals $h_{ij}(t)$ based on numerical iterations of  time reversal. In this peculiar case, we applied 50 iterations of time reversal \cite{Montaldo} in order to diminish the spatio-temporal lobes, followed by an iterative spatial inversion\cite{Lemoult} permitting a better discrimination between sources at the focusing time. The designed filters generate smaller spatial side lobes than the time reversed Green's functions. Undoubtedly one can design more robust signals, especially if the image area grows, and this is only a proof of concept. At this step the system is calibrated and ready for the imaging procedure. 

In the experiment presented in the original paper\cite{metalens}, we applied different weights to each monopolar antenna, placed in the near field of the resonant metalens, and make them emit simultaneously a short pulse. We record with the far field antennas a set of $N=8$ RF signals $s_j(t)$ (This step was done by linearity of the wave equation). These signals supposedly carry the subwavelength information of the object. The image reconstruction starts with cross-correlations between these signals and the filters $h_{ij}(t)$. The final image is the result of this operation at the origin time. The result presented in the paper demonstrated that the Green's functions of two points separated by $\lambda/80$ are sufficiently uncorrelated, and an image resolution of $\lambda/80$ has been achieved. 

Actually, as all of the near fields techniques the object to be imaged disturbs the near field of the resonant metalens, and this why we chose to use a simultaneous emission of small monopoles on the top of ground of plane. Thus, the experimental setup did not permit us to perform 2D imaging because of the original arrangement of the monopoles. But, for these ends we could use a wax prism (and a contrast object placed on the top of it) illuminated in total internal reflection to reach the full 2D superresolved image. For the range of frequency considered here this prism would be very heavy and we will wait to design a smaller lens to perform this experiment. Eventually, the design of the resonant metalens with its high symmetries is not a break to achieve 2D imaging. As stated in the original paper\cite{metalens}, the square geometry gives four distinct radiation directivities for the Bloch modes. Thus, at a given frequency where several Bloch modes are degenerate, the spatial degrees of freedom permit to recover independently the information carried by each of them.

\section{Influence of losses}
Ignoring losses, the limitation for the metalens seems to be the periodicity of the medium: the mode with the highest transverse wavenumber that experiences a resonance has the periodicity of the medium. Thus in the far field region, one can measure information of the object as thin as the lattice parameter. 

Nevertheless, when considering losses, a new limitation appears. Losses deteriorate the resonance $Q$-factor. We have seen that the energy stored in a given mode of the wire medium decreases through radiative decay only when considering perfect conductors. It permitted to express a radiative quality factor and from now we refer to it as $Q_\textrm{rad}$. Introducing losses in metals (and/or in the dielectric structure) results in a degradation of the resonance that we will quantify with a loss $Q$-factor $Q_\textrm{loss}$. The resonance quality factor $Q$ is related to these two quantities by:

\begin{equation}
\frac{1}{Q}=\frac{1}{Q_\textrm{rad}}+\frac{1}{Q_\textrm{loss}}
\end{equation}

From the previous article, we know that $Q_\textrm{rad}$ is roughly proportional to $(DL k_{\perp_x}k_{\perp_y})^2$. The $Q_\textrm{loss}$ can be evaluated in a simple way when considering only the ohmic loss. First, using the fact that the modes are TEM inside the structure, the density of energy along the wire is constant. In the transverse directions we have sine functions, thus the energy stored introduces a $1/2$ constant. Thus, the energy stored in a given mode can be estimated by:

\begin{equation}
w_\textrm{stored}=\frac{D^2}{4}\frac{L}{2}\frac{E_0{}^2}{\mu_0 c^2}
\end{equation}

Second, the surface currents on the conductor surfaces can be estimated from the TEM fields inside the wire medium, and the ohmic losses per unit conductor area are then integrated over the whole area. The surface resistance of lossy metals depends on the skin depth $\delta$ and the metal resistivity $\rho$ and is equal to $\rho/\delta$. Assuming that the magnetic field inside the wire medium evolves as $\cos({\pi z}/{2 L})$, we can finally express the lost energy as:

\begin{equation}
w_\textrm{loss}=\frac{1}{\omega} \frac{\rho}{\delta} \frac{L}{2}(2+\pi)r \frac{1}{\mu_0{}^2 c^2}\sum_\textrm{wires}E_m{}^2
\end{equation}

 \noindent where $E_m$ refers to the amplitude of the electric in the $m$th wire which is linked to the Bloch mode considered. The resonant eigenmodes inside the wire medium correspond to sine functions thus the sum can be approximated by $N^2 E_0{}^2/4$ where $N$ corresponds to the number of wires in one direction. After simplification the $Q_\textrm{loss}$ writes:
 
\begin{equation}
Q_\textrm{loss}=\frac{D^2}{\delta N^2(2+\pi)r}
\end{equation} 
 
  One can notice that $Q_\textrm{loss}$ does not depend on the mode considered (assuming the skin depth to be constant for the range of frequency used) and depends only on the design of the metalens. The impact of losses manifest essentially when $Q_\textrm{rad}$ becomes comparable to the loss constant which occurs for large transverse wavenumbers, and losses tend to decrease the $Q$-factor. This decrease of the resonance quality has two consequences. First, losses have an influence on the amplitude radiated by a given mode. Second, the decrease of the Q-factors results in an increase of the resonance linewidth. 

\subsection{Emission decrease}

In order to illustrate this impact on imaging, we describe a metalens with two parameters: the length of its side $D$ and its number of resonators $N+1$ in this direction (the lattice period $a$ is simply given by $D/N$). The projection of a field source placed at the input onto the eigenmodes write:

\begin{equation}
P(x,y)=\sum\limits_{{m,n}=1}^{N} A_{m,n} \sin\left(\frac{m\pi}{D}x\right)\sin\left(\frac{n\pi}{D}y\right)
\end{equation}   

Each mode experiences its own cavity resonance described by its own Q-factor $Q_{m,n}$ and its own resonance frequency $f_{m,n}$. Then it propagates toward the far field region with an efficiency proportional to the square root of the radiative $Q$ factor. The far field records permit to have a reconstruction of the point source that can be written:

\begin{equation} \label{resolution_eq}
A(x,y) \propto \sum\limits_{{m,n}=1}^{N} A_{m,n} \sqrt{\frac{Q_{m,n}}{Q_{\textrm{rad}_{m,n}}}} \sin\left(\frac{m\pi}{D}x\right)\sin\left(\frac{n\pi}{D}y\right)
\end{equation}

\begin{figure}
\includegraphics[width=\columnwidth]{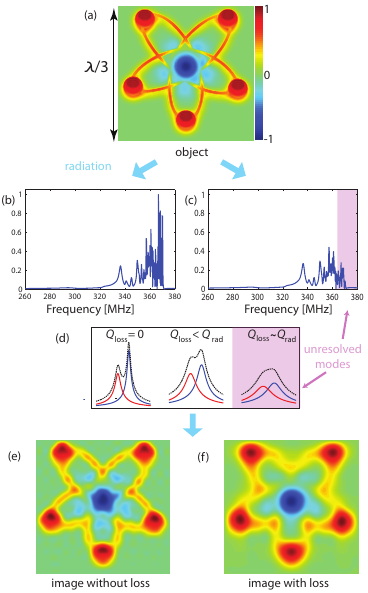}
\caption{\label{resolution} Illustration of the image resolution. (a) A 2 dimensional sub-wavelength field profile corresponding to our object. (b-c) The field spectrum radiated in a given direction by the object placed near the resonant metalens respectively without and with considering losses. (d) Illustration of the impact of losses: the losses tend to decrease the resonant $Q$-factor of a given mode which manifests in a decrease of the resonance amplitude and a decrease in the resonance linewidth. When the resonance linewidth becomes higher than the spectral distance between two consecutive modes they cannot be resolved.  (e-f) According to the criterion we plot the super-resolved image obtained respectively without and with losses in the wires. Without loss, all of the modes can be resolved and the resolution is given by the period lattice, here $\lambda/80$. While in the presence of losses the most sub-wavelength modes are lost, and the resolution decreases down to $\lambda/40$ which is still a deeply sub-wavelength resolution.}
\end{figure}

If there is no loss,  $A(x,y)$ gives the exact projection $P(x,y)$ of the source onto the eigenmodes (figure \ref{resolution}). But, when considering losses, there is a diminution of the amplitude received in the far field due to $Q_\textrm{loss}$. It explains why the focal spot obtained in the time reversal procedure is not as thin as the period lattice. Indeed, time reversal only matches phases of signals and it cannot compensate for the attenuation. For imaging scope, the decrease of $Q$ does not appear dramatic: by applying inversion procedures one can expect to compensate this decrease. This decrease may have an impact when considering noise: if the amplitude radiated becomes lower than the average amplitude of the noise, the inversion procedure will not work. 

\subsection{Resonance linewidth increase}

The influence of the $Q$-factor also manifests in the resonance bandwidth. It results in a mixing of the resonance peaks and it becomes impossible to resolve all of the peaks.  To quantify this aspect, we need the dispersion relation linking the resonance frequency to the Bloch wave number. From this law we can extract the spectral distance $\delta\omega$ between two consecutive modes:

\begin{equation}
\delta\omega=\frac{\partial\omega}{\partial k_\perp}\delta k_\perp
\end{equation} 

\noindent where the spectral distance between two consecutive modes $\delta k_\perp$ has been introduced. Now we have to estimate the group velocity of the medium of interest: this quantity depends on the type of resonators used to build the resonant metalens. In the case of the wire medium, from the theoretical law obtained in the previous article, we can express the group velocity as:

\begin{equation}
\frac{\partial\omega}{\partial k_\perp}=\frac{\omega}{k_\perp}\frac{1}{1+\frac{L}{2}\sqrt{k_\perp{}^2-\left(\frac{\omega}{c}\right)^2}}
\end{equation}

Combining the two previous equations, we estimate the spectral distance (normalized to the resonance frequency) between two consecutive modes, calculated from the dispersion law.  Assuming that the modes are strongly sub-wavelength, the opposite of this quantity writes in the case of the wire medium based resonant metalens:

\begin{equation}
\left(\frac{\delta\omega}{\omega}\right)_{m,n}\approx\frac{2(\delta k_\perp)_{m,n}}{Lk_{m,n}{}^2}
\end{equation} 

\noindent where $(\delta k_\perp)_{m,n}$ corresponds to the $k$-space distance between the mode with transverse wavenumber $k_{m,n}$ and its nearest lower neighbor mode radiating in the same direction. For example, if we consider the radiation in the direction (0$x$), $(\delta k_\perp)_{m,n}=k_{m,n}-\max(k_{m-1,n},k_{m,n-2})$. 

The spectral distance needs to be compared to the resonance linewidth, a quantity contained in the resonance $Q$-factor. We will be able to discriminate two consecutive modes if this spectral length is higher than the resonance linewidth of the $(m,n)$th resonance. Introducing the $Q$-factors already calculated for the wire medium, it gives the following criterion:

\begin{equation}
\frac{2(\delta k_\perp)_{m,n}}{L\pi^2(m^2+n^2)} \geqslant \frac{16}{\pi^3}\left(\frac{1}{Lmn}\right)^2+\frac{1}{\delta N^2(2+\pi)r}
\end{equation} 

When ignoring losses (ie. the skin depth $\delta$ is null) this relation is always true since the left hand operand of the inequality essentially decreases as $1/(n^2+m^2)$ while the right one decreases as $1/(m.n)^2$, and the trueness does not depend on $N$. It means that we can increase $N$ as well as we like in order to achieve the best resolution possible. Without considering losses, the resolution that can be achieved is at best equal to the period of the metalens. As an illustration of this phenomenon, we have performed on figure \ref{resolution} the projection of a two dimensional speculative sub-wavelength source (a) onto the modes of the structure and determinate the image achievable (e). One can notice that the image obtained is not  precisely the original object, but it gives an approximation of the object with a resolution far below the diffraction limit. 

But, when introducing losses, the $Q$-factor characterizing the resonance linewidth takes into account losses and a maximal value for the couple $(m,n)$ appears. Thus, it defines a new resolution limit because the sum in equation (\ref{resolution_eq}) will stop at $(N_\textrm{max},N_\textrm{max})$ instead of $(N,N)$. This implies that the imaging resolution is now limited to $D/N_\textrm{max}$: the resolution has become higher than the lattice parameter.

 The figure \ref{resolution} shows the impact of losses on resolution: the radiated amplitude per mode decreases as well as the resonance linewidth. Thus it becomes impossible to distinguish the resonances near $f_0$ meaning that the modes associated to them are lost.  Finally, the image reconstruction that can be achievable when considering losses with the geometric parameters of our metalens shows a lower resolution than the case without loss, but a sub-wavelength image is still obtained. 

Another interesting remark concerns the design of such a lens.  We have seen that without losses, the smaller the lattice parameter the thinner the resolution: one can argue that it is interesting to increase $N$ as high as possible in order to get the thinnest details. But when considering losses in metals, increasing the number $N$ of wires will drastically increase the ohmic losses. Undoubtedly, an optimal number of resonators for a given transverse dimension exists.

\section{Conclusion}

In this article, we have demonstrated that it is indeed possible to control wave on a scale that is smaller than the wavelength from the far field. As we stated in the previous article, the broadband radiation emanated from the resonant medium contains sub-wavelength information of a source.  Through a time reversal focusing scheme, which has already proved to be a powerful technique dealing with broadband signals, we demonstrated focal spots of width $\lambda/25$. As time reversal cannot compensate for losses, a more accurately method has been investigated for a cross-correlation based imaging procedure. We have performed the proof of concept that an image reconstruction with a resolution of $\lambda/80$ can be achieved. 
Here, we have used half a wavelength long wires as a single resonator, but we would like to emphasize that the concept of resonant metalens should be realizable with any subwavelength resonator, such as split-rings or nanoparticles. We believe that the concept could probably be a good candidate for real time imaging of living cells, with a resolution far better than classical microscopes.

\bibliography{biblioPRB}

\end{document}